\documentclass[10pt,a4paper,conference]{IEEEtran}
\ifCLASSINFOpdf
  \usepackage[pdftex]{graphicx}
  \usepackage{caption}
  \usepackage{subcaption}
  \usepackage{color}
\else
  \usepackage[dvips]{graphicx}
\fi
\begin{document}
%
\title{Deploying SDN in G\'EANT production network*}


%

\author{
\IEEEauthorblockN{
    Pier Luigi Ventre\IEEEauthorrefmark{1}, 
    Jordi Ortiz\IEEEauthorrefmark{2},
    Alaitz Mendiola\IEEEauthorrefmark{3}, 
    Carolina Fern\'andez\IEEEauthorrefmark{4},
    Adam Pavlidis\IEEEauthorrefmark{5},
    Pankaj Sharma\IEEEauthorrefmark{6},
    \\
    Sebastiano Buscaglione\IEEEauthorrefmark{7},
    Kostas Stamos\IEEEauthorrefmark{8}, 
    Afrodite Sevasti\IEEEauthorrefmark{8},
    David Whittaker\IEEEauthorrefmark{9}
}
\IEEEauthorblockA{\IEEEauthorrefmark{1}University of Rome Tor Vergata, Italy,
\IEEEauthorrefmark{2}University of Murcia, Spain,
\IEEEauthorrefmark{3}University of the Basque Country, Spain,
\\
\IEEEauthorrefmark{4}I2CAT, Spain,
\IEEEauthorrefmark{5}NETMODE Laboratory, NTUA, Greece,
\IEEEauthorrefmark{6}RENATER, France,
\IEEEauthorrefmark{7}G\'EANT, Europe,
\\
\IEEEauthorrefmark{8}GRNET, Greece,
\IEEEauthorrefmark{9}Corsa Technologies, Canada
}

}


\maketitle

\begin{abstract}

Since the demand for more bandwidth, agile infrastructures and services grows, it becomes challenging for Service Providers like G\'EANT to manage the proprietary underlay, while keeping costs low. In such a scenario, Software Defined Networking (SDN), open hardware and open source software prove to be key components to address those challenges. After one year of development, SDX-L2 and BoD, the SDN-ization of the G\'EANT Open and Bandwidth on Demand (BoD) services, have been brought to the pilot status and G\'EANT is now testing the outcomes on its operational network. In this demonstration, we show BoD and SDX-L2 "going live" at the G\'EANT production infrastructure. The pilots run on the same underlay infrastructure thanks to the virtualization capabilities of the network devices. Provisioning of the services is covered during the demo. In the final steps of the demonstration, we show how the infrastructure is able to automatically manage network events and how it remains operational in the case of fault events.

\end{abstract}

%
\IEEEpeerreviewmaketitle

\let\svthefootnote\thefootnote
\let\thefootnote\relax\footnotetext{*This work was partly funded by the EU G\'EANT (GN4-2) project \cite{geant} as part of the G\'EANT 2020 Framework Partnership Agreement (FPA) under Grant Agreement No. 731122\\\\}\let\thefootnote\svthefootnote

\vspace{0.5ex}

\vspace{-1.65ex}

\section{Introduction}
G\'EANT \cite{geant}, the 500 Gb/s pan-European provider interconnecting 38 national research and educational networks for a total of 50 million users, is transitioning from doing traditional service provisioning to software-oriented workflows, more dynamic and cost-less, leveraging multi-point software defined fabrics. The current G\'EANT infrastructure is based on traditional control plane architectures (IP/MPLS) running on top of closed and expensive equipment and management tools. The service provisioning includes manual and error-prone procedures. Finally, the enhancements in networking equipment and speed increase have not been accompanied by the same innovation level in the development of new network services. In such a scenario, Software Defined Networking \cite{sdn} (SDN), open hardware and open source software are considered key enabler technologies for networks. In the last years, as part of the GN4 project \cite{geant}, G\'EANT has funded work oriented to apply such technologies in order to update already existing services as well as developing new ones. Examples of this transition include G\'EANT Open \cite{geant_service}, G\'EANT Bandwidth on Demand \cite{geant_service} (BoD), and G\'EANT Testbed Service \cite{geant_service} (GTS). The first is a neutral, policy-free connectivity service between different Research and Education Networks, users and commercial operators. G\'EANT BoD provides multi-domain connectivity services with bandwidth guarantees for a specified duration to the users. The inter-domain coordination is achieved through the Network Service Interface (NSI) Connection Service \cite{nsi}. Lastly, GTS is a continental facility offering geographical virtual testbeds to the research community comprising VMs, bare-metal server, and SDN devices.

\begin{figure*}[t]
\centering
\begin{subfigure}{0.38\textwidth}
  \centering
  \includegraphics[width=1.0\textwidth]{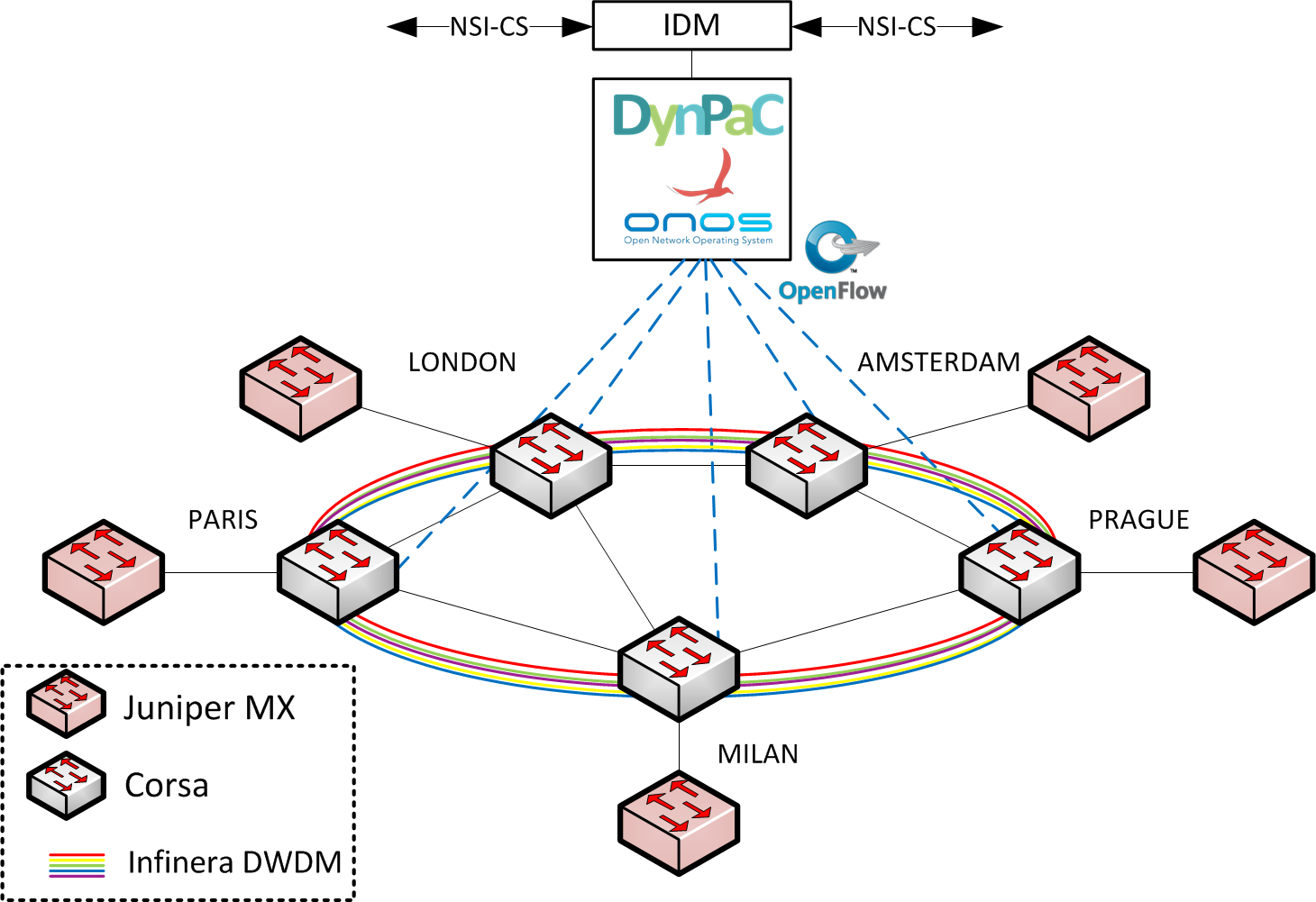}
  \vspace{-4ex}
  \caption{BoD Pilot}
  \label{fig:bod_pilot}
\end{subfigure}
\begin{subfigure}{0.43\textwidth}
  \centering
  \includegraphics[width=1.0\textwidth]{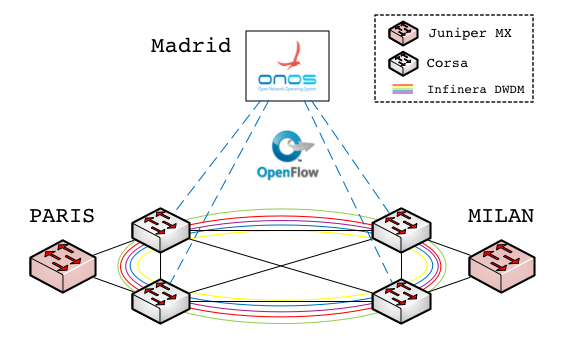}
  \vspace{-4ex}
  \caption{SDX-L2 Pilot}
  \label{fig:sdxl2_pilot}
\end{subfigure}
\vspace{-1.5ex}
\caption{Pilot Deployments}
\label{fig:pilots}
\vspace{-4ex}
\end{figure*}

The demonstration is intended for: i) showing in operation the SDN counterpart of the first two use cases, namely SDX-L2 and BoD, currently running on the G\'EANT SDN infrastructure (Fig. \ref{fig:pilots}); and finally ii) demonstrating how the pan-European operator can test in parallel both services sharing the underlying infrastructure with the GTS.

\section{Deploying in G\'EANT production network}
According to the migration strategy we have defined in \cite{geant_sdx}, the SDN-enabled networking equipment replaces the traditional devices and the services provided by G\'EANT have been re-designed as SDN applications running on top of an SDN controller. The Open Networking Operating System (ONOS) \cite{onos} has been chosen as SDN control platform. At the time of writing, G\'EANT engineers are testing the outcome of this phase on the pilot infrastructure represented in Fig. \ref{fig:pilots}. 

BoD \cite{bod} is intended to evolve the current G\'EANT's BoD, based on Juniper equipment and Junos Space NMS, into an SDN-enabled service. The DynPaC framework is responsible for scheduling services over time and computing the paths to dynamically allocate the resources in the network enforcing the requested service. DynPaC runs on top of ONOS providing with users CLI, GUI and REST interfaces. Through the latter, the system is able to speak with the Inter-Domain Manager (IDM) and with the AutoBAHN web portal. The first understands the NSI protocol enabling multi-domain services. While the second is the interface for the legacy G\'EANT's BoD service. Therefore, BoD enables the coexistence of SDN islands with legacy service. VLAN tagging is used to stitch different technological domains. The SDN service also provides link failure recovery capabilities transparently and automatically managed by the ONOS application. SDX-L2 \cite{geant_sdx} is the SDN-inization of the G\'EANT Open service. It is an ONOS application that allows the automated provisioning of layer 2 tunnels between endpoints, which can be plain Ethernet or VLAN interfaces. The network operator can manage and monitor the services through a CLI and a GUI, which accept high-level customer requests and the application translates them into ONOS point-to-point Intents. SDX-L2 can run as a single instance or in the form of cluster, composed by multiple (3, 5, 7 or more) instances, all functionally identical to each other. Failures in the control plane are managed through the redundancy of the cluster. Data plane failures are automatically resolved through transparent rerouting mechanisms.

As regards the network devices, the Corsa's DP2000 family \cite{corsa} has been deployed in the G\'EANT infrastructure, which offers three product variants, each providing a variety of 1/10G SFP+ and 100G QSFP28 interfaces as well as scalability through stacking up to 2.4T. The platform supports full virtualization of the hardware and the devices' resources can be used to build up to 256 virtual SDN switches or routers, the so-called Virtual Forwarding Contexts (VFCs). The logical interfaces of the VFCs can correspond to a physical interface or an encapsulated VLAN tunnel. Leveraging such constructs we are able to run both pilots on the same infrastructure achieving improved bandwidth management and better traffic isolation with respect of other virtualization technologies. OpenFlow 1.3 is used as southbound interface. Specific drivers for ONOS have been developed to map the device-agnostic forwarding instructions of the applications to specific match-action in the available OpenFlow tables of the Corsa pipelines. These drivers leverage additional features of the Corsa products which guarantee even more efficiency and data-plane scale.

\section{Demonstration}
Fig. \ref{fig:pilots} shows the pilot deployments which extends over 6 PoPs of the G\'EANT infrastructure. SDX-L2 and BoD use respectively 4 and 5 VFC instances, which are distributed over Milan, London, Prague, Amsterdam and Paris locations. The virtual switches are connected using redundant links at 10G provided on top of G\'EANT's DWDM. Juniper MX switches connect the Pilot to G\'EANT's legacy infrastructure as well as some clients, at least one per PoP. The clients are connected using 1G links. Madrid hosts the control plane. For SDX-L2, a cluster of 3 ONOS instances controls the SDN fabric. As regards BoD, 3 VMs serving as ONOS controller containing the DynPAC app, the Inter Domain Manager (IDM) and Autobahn Portal. The BoD demo and SDX-L2 demo will take place on different tenant networks over the same underlay which is already running the GTS virtual testbeds. This has been made possible thanks to the full virtualization capabilities of the Corsa devices. 

At the beginning, the BoD demonstration will show how services can be requested providing start and end times and how they are instantiated and deployed automatically in the network. The services will be requested by means of CLI, ONOS' GUI and the AutoBAHN portal. With the services in place, we will show bandwidth enforcement through OpenFlow meters, VLAN translation and link failure recovery in SDN, as well as coordination with other BoD legacy domains through the NSI interface. A link failure is simulated in the pilot by detaching the logical tunnel between the physical Corsa port with a VFC OpenFlow port. This action is perceived by ONOS as the loss of a port and in turn a link loss between two VFCs triggering a recovery procedure in DynPaC. As regards SDX-L2, the demonstration will show the functionalities of the pilot represented in Fig. \ref{fig:pilots}. The layer 2 circuits are created leveraging the SDX-L2 application running on top of ONOS and connectivity is demonstrated. Then, failure recovery is showed simulating a link failure: all affected circuits are automatically rerouted with a minimal disruption for the users. SDX-L2 has been designed to be a distributed application. Thus, after a failure of an ONOS instance, the application keeps working without any effect on the infrastructure and on the active services. At the end of the demonstration, the developed CLI and GUI are showcased to the audience.

\vspace{-1.0ex}






%

\bibliographystyle{IEEEtran}
\bibliography{main}

\end{document}